\documentclass{PoS}

\newcommand{\avg}[1]{{\langle \hspace{0.2em} #1 \hspace{0.2em} \rangle}}
\newcommand{\overtilde}[1]{\stackrel{\sim}{#1}}

\title{Universal properties of the confining string in the random percolation model}

\ShortTitle{Universal properties of the confining string in the random percolation
model}

\author{Pietro Giudice, Ferdinando Gliozzi, \speaker{Stefano Lottini}\\ %
        Universit\`a di Torino \& INFN sez. di Torino\\
        E-mail: \email{giudice, gliozzi, lottini@to.infn.it}}

\abstract{Random percolation can be fully interpreted as a confining pure gauge theory.
With numerical high-precision measurements of Polyakov-Polyakov correlators at finite
temperature, we could well observe the presence of shape effects due to rough fluctuations
of the confining string, in complete agreement with the universality predictions 
up to the next-to-leading order.}

\FullConference{The XXV International Symposium on Lattice Field Theory\\
                 July 30 - August 4 2007\\
                 Regensburg, Germany}

\begin{document}

\section{Introduction}
	It is now widely believed that a $D$-dimensional confining gauge theory can be described, in the infrared limit, by a suitably chosen effective string theory. If the interquark separation is large, indeed, the relevant degrees of freedom are the oscillations of the confining string worldsheet in the $D-2$ transverse directions to the loop plane.
	
	The Nambu-Goto action, which is proportional to the string worldsheet area, is however consistent under Lorentz covariance only in dimensionality $D=26$; by slightly modifying the action with the addition of a nonpolynomial term, one gets the action \cite{ps}
	\begin{equation}
			S = \sigma \int \mathrm{d}^2 \xi \big\{ 
			\partial_+ X_\mu \partial_- X^\mu + \beta \frac{(\partial^2_+ X_\nu \partial_- X^\nu)
			(\partial_+ X_\nu \partial^2_- X^\nu)}{(\partial_+ X_\mu \partial_- X^\mu)^2}
			+ \mathcal{O}(\partial^{-3})
		\big\} \, ,
	\end{equation}
which fixes the above problem and, while reproducing the Nambu-Goto features, gives physical predictions in any $D \geq 3$. With the help of open-closed string duality in $D=3$ \cite{lw}, and from some exact calculations available in any dimension, it has been argued that these effective theories are, to some finite order, universal.

	The validity of the effective string picture for confinement is strongly related to the phenomenon of \emph{roughening}: it is exactly when the loop surface can undergo quantum fluctuations on any length scale that the identification with a bosonic, massless string is possible.

	By applying the Riemann zeta function regularisation prescription (to the Nambu-Goto effective theory), it is possible to work out the functional form of observables such as Wilson loops in the form of an expansion in the small quantity $1/(\sigma \cdot A)$, with $\sigma$ string tension and $A$ the minimal area spanned by the loop contour \cite{df}.

	This work aims at determining the validity of the universality prediction by comparison with Monte Carlo data collected in a particular gauge theory, namely the random percolation model. In the following Sections, the observable of interest is presented in more detail, then a brief description of the gauge theory is given; afterwards, we describe the technique we used to obtain the data, and the results and conclusions that we drew.

\section{The Polyakov-Polyakov correlation function}
	We focused on the behaviour of the Polyakov-Polyakov correlation function at finite temperature in a $(2+1)$-D system. That is, the lattice is a $L_x \times L_y \times L$ slice, with $L_x$ and $L_y$ large enough to represent the spatial extent and $L = \frac{1}{T}$ the inverse temperature. We considered a couple of Polyakov loops orthogonal to the spatial direction and at a distance of $R$ lattice spacings; the (connected) correlation function in this case is denoted with $\avg{P(0)P^*(R)}$.
	
	Since the subleading terms in the correlator are a shape effect rather than exhibit a size dependence, we introduce the aspect ratio of the cylinder bounded by the loops and the sides of the lattice:
	\begin{equation}
		\tau \equiv \frac{i L}{2 R} \quad ; \qquad q\equiv e^{2 \pi i \tau} \, .
	\end{equation}
	
	The Polyakov-Polyakov correlator is then expected to follow the next-to-leading order (NLO) prediction
	\begin{equation}
		\label{eq:pp_nlo}
		<P(0)P^*(R)> = \frac{e^{-cL-\sigma RL-\frac{(D-2)\pi^2L
				[2E_4(\tau)-E_2^2(\tau)]}{1152\sigma R^3}+\mathcal{O}(1/R^5)}}{\eta(\tau)^{D-2}} \, ,
	\end{equation}
	where the functions $\eta$ (Dedekind eta), $E_2$ and $E_4$ (second and fourth Eisenstein functions) are defined by:
	\begin{eqnarray}
		\eta(\tau) & \equiv & q^{\frac{1}{24}} \prod_{n=1}^\infty (1-q^n) \, , \\
		E_2(\tau)  & \equiv & 1- 24 \sum_{n=1}^\infty \sigma_1(n) q^n \, , \\
		E_4(\tau)  & \equiv & 1+240 \sum_{n=1}^\infty \sigma_3(n) q^n \, ;
	\end{eqnarray}
	the functions $\sigma_i(n)$ here represent the sum of the $i$-th powers of all divisors of $n$. This functional form for $<P(0)P^*(R)>$ is expected to be universal, to this order, for a wide variety of string functionals in the Nambu-Goto family \cite{dr}.
	
	A direct numerical observation of these extremely fine corrections is, in most lattice gauge theories, unfathomably beyond any computational possibility.	However, by choosing a particularly simple gauge model, a better numerical accuracy can be reached and the prediction can be adequately tested. This has been done on the three-dimensional $Z_2$ gauge model in \cite{chp}.

\section{The random percolation model}
	In this work, the model we chose as laboratory is the random percolation model. By suitably defining the observables, indeed, it can be shown that this theory, although remaining somewhat peculiar, behaves exactly as expected in a ordinary confining pure gauge theory \cite{glpr}. This interpretation is supported by a number of theoretical arguments (such as the center vortex picture for confinement and the reformulation of the dual gauge system in terms of Fortuin-Kasteleyn clusters) as well as many numerical evidences concerning physical expectations (universal ratios, string tension scaling, glueball spectrum and so on). Nevertheless, the percolation model is somewhat particular, having a trivial partition function and gauge group ($Z \equiv 1$ and $G = \{ e \}$); indeed it can be thought of as the $q\to 1$ limit of the $q$-state Potts model.

	The important aspect is that, as will be shown in the next Sections, even though the explicit formulation of the model does not involve strings at all, clear signals of a rough string behaviour are identified. It is already known that in this model a loop obeys the predicted functional form at least at the leading-order (LO); the question we address here is whether the model confirms the expected universality even at the NLO.

	In the (bond-)percolation model, each link of the lattice is independently set to \emph{on} or \emph{off} according to some fixed probability $p$, which plays the role of a coupling constant. Given a lattice geometry, there exist a critical value $p_c$, corresponding to the sudden appearance of an infinite connected cluster; this is a second-order transition point and will represent the deconfinement transition, which can be easily mapped to a finite-temperature transition at $T=T_c$ because the critical threshold depends on the temperature (thought as the inverse extension of the lattice in the periodic direction).

	The starting point in this pure gauge framework is the definition of loop observables: given a loop with contour $\gamma$, the value of the associated Wilson loop (or any other object, such as a couple of Polyakov loops) is defined as:
	\begin{eqnarray}
		W(\gamma) = 1 \quad & \Longleftrightarrow & \quad \mbox{\textit{no cluster is topologically linked to the contour 
		$\gamma$}} \, ; \\
		W(\gamma) = 0 \quad & \Longleftrightarrow & \quad \mbox{\textit{otherwise}} \, . 
	\end{eqnarray}
	From this definition follows naturally an area/perimeter law for large enough loops, which distinguishes between the confined ($p>p_c$) and deconfined ($p<p_c$) phases of the theory, providing in the former a well-defined string tension and $p$-dependent critical temperature that scale according to
	\begin{eqnarray}
		\label{eq:scalings}
		\sigma & = & S (p-p_c)^{2\nu} \, ,\\
		T_c & = & T_0(p-p_c)^\nu \, .
	\end{eqnarray}
	
	The value of a loop, moreover, is insensitive to changes in the configuration that do not alter the loop structure; this is a sort of gauge invariance of the theory, besides drastically reducing the computational effort required for numerical investigations.

\section{Methodology}

	We worked on a cubic lattice of size $128^2 \times L$, where $L$ is the periodic inverse temperature. For each choice of the occupation probability $\overtilde{p}$, corresponding to some deconfinement temperature $T_c = 1/L_c$, we took systems at various temperatures in the range $\frac{T_c}{2} \stackrel{<}{\sim} T \stackrel{<}{\sim} T_c$.
	
	For each of these systems, we measured $\avg{P(0)P(R)}$ (the dagger can be dropped, since this theory deals only with real loop values), varying the distance between the two Polyakov lines from $R_0$ to $R_{\mathrm{max}}$; the data are unbiased by the spatial finiteness of the system up to at least half of the system size, and we safely chose $R_0=8, R_{\mathrm{max}} = 50$.
	
	The expectation value of a Polyakov loop couple is defined in terms of topological linking with the rectangle (periodic in one direction) which has the loops as boundary; once enough numerical accuracy is reached, then, it is possible to fit the data sets to Eq.~\ref{eq:pp_nlo}, extracting the string tension by looking at a plateau in the choice of the fit interval.
	
	As for the dependence of the string tension from the temperature $T=1/L$, from the same Eq.~\ref{eq:pp_nlo} it is possible to find that, for asymptotically large $R$,
	\begin{equation}
		\label{eq:sigmal}
		\sigma(L) = \sigma - \frac{\pi}{6 L^2} - \frac{\pi^2}{72 \sigma L^4} + \mathcal{O}(1/L^6) \qquad ; \quad L = \frac{1}{T} \; , \; R \to \infty \, ;
	\end{equation}
	here, the symbol $\sigma(L)$ denotes the physical quantity that scales well with the temperature, while $\sigma$ is only a parameter in the fit that will be attempted to check this temperature scaling. $\sigma$ would represent the zero-temperature string tension (thought of as a function of $\overtilde{p}$) if the above NLO formula were exact.

	We chose to focus our attention on two values of $\overtilde{p}$: at $\overtilde{p}_1 = 0.272380$ (critical for $1/T = 6$), we considered the temperatures in the range $1/T = 7,\ldots,15$, while at $\overtilde{p}_2 = 0.268459$ (critical for $1/T = 7$), we examined $1/T = 8,\ldots,15$.\footnote{Note that the zero-temperature critical point is located at $p_c(0)\simeq 0.248812$\,.}
	
	To reach an acceptable statistics, we collected data from $10^5$ configurations for each value of $\overtilde{p}, L$. For different blocks of 8 values of $R$ we used independently-generated configurations.

	There is another signal that the theory is in the rough phase: the quantity
	\begin{equation}
		\label{eq:ft}
		f(t) = \frac{\sigma(T)}{T^2_c} \;\; ; \;\; t \equiv \frac{T_c - T}{T_c}
	\end{equation}
	should be a universal ratio with no adjustable parameters, that is, it is supposed not to vary for different realisations of the transition (meaning different choices of $\overtilde{p}$ and the corresponding $T_c$).
	
\subsection{Algorithm}
	Due to the particular nature of the random percolation model, each configuration can be generated independently from scratch, by simply filling an empty lattice with links that are randomly switched on. The tricky part is the measurement of the topological linking of the resulting cloud with a given surface; to this end, the first thing to do is to ``clean up'' the configuration, getting rid of dead ends and simply-connecting bridges between loop structures. This is done once for the whole configuration.
	
	On this ``minimal gauge'' configuration, then, the loop is measured in all possible spatial positions with the technique of reconstructing each time the clusters in the configuration (by means of the Hoshen-Kopelmann algorithm) keeping track of the crossings of the loop surface, to detect nonzero winding numbers.

\section{Data and results}
\begin{figure}
	\begin{center}
		\includegraphics[scale=0.23,angle=270]{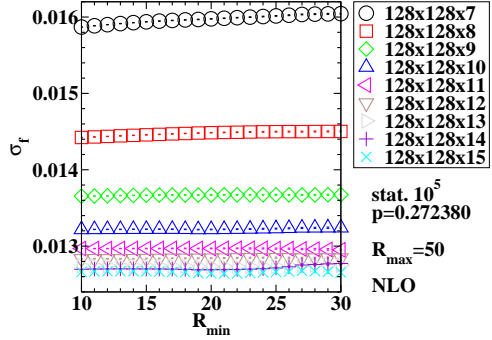}
	\end{center}
		\caption{Zero-temperature string tensions $\sigma$ obtained with the NLO formula at $\overtilde{p}_1$ at different temperatures. The quantity $R_\mathrm{min}$ denotes the lower end of the fit interval, the upper one being fixed to 50. Note that, for large enough values of $L$, a plateau appears quite soon.}
		\label{fig:nlofits}
	\end{figure}

	We tried to fit the measured correlation functions to Eq.~\ref{eq:pp_nlo}: for temperatures far enough from the deconfinement point ($L$ greater than 9 for $\overtilde{p}_1$ and 8 for $\overtilde{p}_2$), and large enough distances ($R\geq 8$ and $R\geq 9$ respectively), the NLO formula works well and gives reliable plateau for the string tensions $\sigma$ (Fig.~\ref{fig:nlofits}).
	
\begin{figure}
	\begin{center}
		\includegraphics[scale=0.16,angle=270]{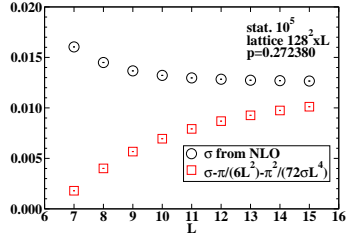}
	\end{center}
		\caption{Temperature dependence, for $\overtilde{p}_1$, of the physical string tension $\sigma(T)$ (squares) and of the fit parameter $\sigma$ (circles), from which the former is extracted.}
		\label{fig:sigmal}
	\end{figure}

	The quantities $\sigma$ thus obtained can be inserted in the asymptotic formula (\ref{eq:sigmal}) to reconstruct the physical quantity $\sigma(T)=\sigma(L)$. However, since the expansion is truncated at $\mathcal{O}(L^{-4})$, the precision of the data allows to observe a residual dependence on $L$ (Fig.~\ref{fig:sigmal}).

	One can then look for the next term in the corrections, keeping in mind that the predicted universality will not be valid any more beyond the NLO. So, we made, for this first model-dependent term, the Ansatz
	\begin{equation}
		\sigma(L) = \sigma - \frac{\pi}{6 L^2} - \frac{\pi^2}{72 \sigma L^4} + \frac{\pi^3}{C \sigma^2 L^6} + \mathcal{O}(1/L^8) \, ,
	\end{equation}
	in which a new parameter, $C$, has appeared.
	
	By fitting the data sets with this functional form, we could well identify stable values both for the ``true'' zero-temperature string tension $\sigma(T=0;\overtilde{p})$ and the coefficient $C$ in the $L^{-6}$ term (moreover, the two results for $C$ are compatible, as one would have hoped):
	\begin{footnotesize}
	\begin{center}
		\begin{tabular}{|r|r|r|}
		\hline
			$\overtilde{p}$ & $\sigma(T=0)$ & $C$ \\
		\hline
		\hline
			$\overtilde{p}_1$ & $0.0126(1)$ & $296 \pm 5$ \\
		\hline
			$\overtilde{p}_2$ & $0.00925(2)$ & $302 \pm 4$ \\
		\hline
		\end{tabular}
	\end{center}
	\end{footnotesize}
	
\begin{figure}
	\begin{center}
		\includegraphics[scale=0.21,angle=270]{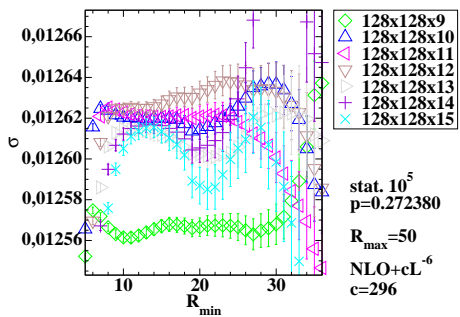}
	\end{center}
		\caption{String tensions plateaux obtained with the $\mathcal{O}(L^{-6})$ formula for $\avg{P(0)P(R)}$, for the data at $\overtilde{p}_1$. Now the position of the plateau is temperature-independent.}
		\label{fig:fit2}
	\end{figure}

		\vspace{-0.2cm}As a circular check, we inserted this non-universal correction in Eq.~\ref{eq:pp_nlo}, and re-fitted the data sets: the plateaux for different temperatures (as long as they are not too close to the critical point) now coincide, confirming the estimates for $\sigma(\overtilde{p})$ (Fig.~\ref{fig:fit2}).

\begin{figure}
	\begin{center}
		\includegraphics[scale=0.19,angle=270]{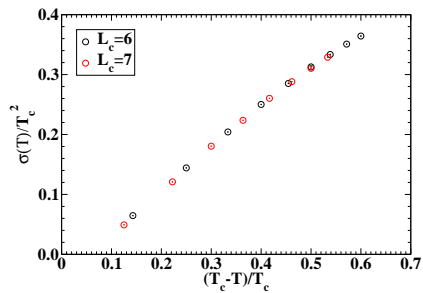}
	\end{center}
		\caption{Behaviour of the universal ratio $f(t)$ for the two examined values of $\overtilde{p}$. Note that the function does not fall to zero exactly at $T=T_c$.}
		\label{fig:univ}
	\end{figure}

	We could also confirm that the adimensional ratio $f(t)$ of Eq.~\ref{eq:ft}, as a function of the reduced temperature, indeed does not depend of the choice of $\overtilde{p}$, as expected in any confining theory in the rough phase (Fig.~\ref{fig:univ}).

\section{Conclusions}

	In this work the universality of the string behaviour up to the NLO has been numerically proven in a particular realisation of a confining gauge system in the rough phase. In the percolation model we have confirmed that the Polyakov-Ployakov correlation function follows the expected behaviour; moreover, we could identify the first model-dependent correction in the behaviour of $\sigma(L)$ with a stable coefficient\footnote{The coefficient $C$ appears compatible with the integer 300, suggesting it comes, as expected, from some multiplicity count. However, by lack of information about the small-scale behaviour of the theory, we could not delve deeper into this issue.}.

	The universality of $f(t)$ confirms the validity of the string picture in this model (for large enough interquark separations). However, the function $f(t)$ does not seem to drop to zero at exactly $T=T_c$. This should be due to the fact that, approaching criticality, the algorithm used has to include all topological classes of configurations. A further analysis on this aspect could be carried on.

\end{document}